\documentclass[aps,prl,preprint,showpacs,preprintnumbers]{revtex4}
\usepackage{amssymb,latexsym,amsfonts,amsmath,color}
\usepackage{framed}
\usepackage{natbib}
\newcommand{\rem}[1]{}

\newcommand{\pa}{\partial}

\newcommand{\qed}{{\vrule width8pt height8pt}}

\rem{

}

\newtheorem{thm}{Theorem}
\newtheorem{dfn}{Definition}

\newtheorem{remark}{Remark}

\newcommand{\spair}[2]{{\langle {#1}\, , \, {#2}\rangle}}
\newcommand{\pair}[2]{{\big\langle {#1}\, , \, {#2}\big\rangle}}

\begin{document}

\title{Euler's fluid equations: Optimal Control vs Optimization}
\author{Darryl D. Holm}
\affiliation{Department of Mathematics, Imperial College London, SW7 2AZ, UK}




\date{\today}
\begin{abstract} \noindent
An optimization method used in image-processing (metamorphosis) is found to imply Euler's equations for  incompressible flow of an inviscid fluid, without requiring that the Lagrangian particle labels exactly follow the flow lines of the Eulerian velocity vector field. Thus, an optimal control problem and an optimization problem for incompressible ideal fluid flow both yield the \emph {same} Euler fluid equations, although their Lagrangian parcel dynamics are \emph{different}. This is a result of the \emph{gauge freedom} in the definition of the fluid pressure for an incompressible flow, in combination with the symmetry of fluid dynamics under relabeling of their Lagrangian coordinates.  Similar ideas are also illustrated for $SO(N)$ rigid body motion.

\rem{This reformulation of Euler's fluid equations by using the metamorphosis method from image-processing provides a framework for optimal fluid flows in which extremizing the total cost function yields a true minimum.}

\bigskip

\end{abstract}

\maketitle




\rem{%
\newcounter{thm}

\newcounter{remark}
\newcounter{dfn}
\setcounter{thm}{0}
\setcounter{remark}{0}
\setcounter{dfn}{0}

\refstepcounter{thm}
\refstepcounter{remark}
\refstepcounter{dfn}
}%

\noindent

Hamilton's principle for ideal fluid flow might be summarized by saying that \emph{water moves as well as possible to get out of its own way} \cite{JM1990}. This phrase could make the challenge of optimal control of fluids seem daunting, particularly when combined with Le Chatelier's principle that any complex system will respond to resist the effects of an external force.  The question pursued here is whether Euler's fluid equations (EFE) represent optimal control, or only optimization. As it turns out, the geodesic flow represented by the EFE is found to arise from either formulation. 

\begin{dfn}[Optimal control]$\quad$\\
\rm
Optimal control problems consist of \cite{AgSa2004}:
\begin{itemize}
\item A differentiable manifold $M$ on which state variables $Q\in M$ evolve in time $t$ during an interval ${I}=[0,T]$ along a curve $Q: {I} \to M$  from $Q(0)=Q_0$ to $Q(T)=Q_T$, with specified values $Q_0,Q_T\in M$;
\item A vector space $V$ of control variables ${U}\in V$ whose time dependence ${U}: I\to V$ is at our disposal to affect the evolution  $Q(t)$ of the state variables;
\item A system of first-order evolutionary partial differential equations (state equations) defined on the tangent bundle $TM$ by the vector field $F:M\times V\to TM$, 
\begin{equation}
\dot{Q}=F(Q, {U}),
\label{statesyst-eqn}
\end{equation}
and introduced as a \emph{constraint} that relates the unknown state and control variables $(Q(t), {U}(t)): I \to M\times V$,
and
\item A cost functional depending on the state and control variables
\begin{equation}
S:=\int_0^T \ell(Q, {U})\,dt,
\label{cost-optimconprob}
\end{equation}
whose minimization is the goal, subject to the prescribed initial and final conditions at $Q(0)$ and $Q(T)$, and the state equations (\ref{statesyst-eqn}). The integrand $L: M\times V\to \mathbb{R}$ is called the Lagrangian, which is assumed to be  continuous and continuously differentiable on $M\times V$.

\end{itemize}
\end{dfn}
Thus, solving a standard optimal control problem requires finding time-dependent state and control variables $(Q(t), {U}(t))$ that optimize a given cost functional $S$ subject to the exact enforcement of an evolutionary system of state equations (\ref{statesyst-eqn}), while satisfying prescribed initial and final conditions, $Q_0$ and $Q_T$. 

The optimal control  problem is traditionally formulated by introducing a pairing $\spair{\,\cdot\,}{\,\cdot\,}:TM^*\times TM\to \mathbb{R}$ between the tangent space of state variables $(Q,\dot{Q})\in TM$ and the cotangent space of \emph{costate} $(Q,P)\in T^*M$ variables. In terms of this pairing, the state equation may be  enforced as a constraint on variations of the cost function, by using the classical method of Lagrange multipliers. Thus, the optimal control problem in this formulation becomes, 
\begin{equation}
\min_{Q(t),U(t)}\int_0^T \Big[ \ell(Q,U)
+
\pair{P}{\dot Q-F(Q,U)} \Big] dt\,,
\label{opticonprob-Clebsch}
\end{equation}
for which standard methods of variational calculus are available. (The variations are to be taken at fixed values of the endpoints $Q_0$ and $Q_T$.)

\begin{dfn}[Optimization by metamorphosis]$\quad$\\
\rm
The term \emph{metamorphosis} refers to a class of optimization methods used for performing image registration by finding the optimal flow along a curve in the group of of diffeomorphisms ${\rm Diff}(M)$ (smooth invertible maps with smooth inverses) acting on a differentiable manifold $M$ of image properties (states) $Q\in M$ defined over a given spatial domain $\cal D$ \cite{TrYo2005}. Such an optimal flow is sought as a geodesic time-dependent curve with respect to a certain metric on the tangent space $T{\rm Diff}(M)$. Hence, one chooses control variables $U\in V = T{\rm Diff}(M)$. One version of optimization by metamorphosis replaces the optimal control problem (\ref{opticonprob-Clebsch}) by the minimization of a sum of norms integrated over the control time interval. This version of metamorphosis may be expressed as the following optimization problem, 
\begin{equation}
\min_{Q(t),U(t)}\int_0^T \Big[ \ell(Q,U)+\frac{1}{\sigma^2}\|\dot Q-F(Q,U)\|^2 \Big] dt\,,
\label{optimz-prob-metam}
\end{equation}
for a positive real constant $\sigma^2 $ and a chosen norm $\|\,\cdot\,\|: T{\rm Diff}(M)\to\mathbb{R}$,  while satisfying prescribed initial and final conditions, $Q_0$ and $Q_T$.

The optimization problem (\ref{optimz-prob-metam}) is solved by finding controls $U(t)$ that steer the state variable $Q$ along a time-dependent curve $Q(t)$ leading from $Q_0$ to $Q_T$, obtained by minimizing the cost   of applying the controls $S$ in (\ref{cost-optimconprob}), while enforcing the state equations (\ref{statesyst-eqn}) within a certain tolerance $\sigma$. The solutions $(Q(t),U(t))$ of an optimal control problem and its corresponding problem of optimization by metamorphosis need not coincide, even when their cost functions and state equations are the same. 

Optimization by metamorphosis finds applications in problems of image registration and recognition of patterns, e.g., in the analysis of medical images obtained using MRI, CT and other imaging technologies  \cite{TrYo2005,HoTrYo2009}. In these applications, shapes must be matched, or at least compared to one another. The matching or comparison procedure is formulated as an optimization problem whose goal is to minimize the sum of the chosen norm on the tangent space of the path plus another norm associated with the estimated error, or tolerance, of the measurement process. This formulation of optimization by \emph{metamorphosis} has the advantage of introducing a Riemannian structure that ensures that the extremals in the distance between the images are genuine minima, rather than being saddles. For more discussion of the metamorphosis approach, see  \cite{TrYo2005,HoTrYo2009}.
\end{dfn} 

In addition to image analysis, an interest in controlling diffeomorphisms also arises in ideal fluid flows.
When applied to fluid dynamics, optimization by metamorphosis introduces a penalty defined by a metric on the tangent space of the \emph{inverse} flow (also known as the back-to-labels map for fluids).  The penalty introduces an additional cost in the kinetic energy of labels whose paths deviate rapidly from the Eulerian characteristics following the \emph{forward} flow. However, the metamorphosis approach does not \emph{constrain} the fluid labels to follow exactly along the forward  flow lines. \medskip

{\bf Objective} 
The present paper shows that optimization by metamorphosis recovers the classical Euler fluid equations (EFE), even though the metamorphosis approach does not require that the fluid labels exactly follow the characteristic curves of the Eulerian fluid velocity. This result implies that the standard Lagrangian representation of fluid dynamics as labelled fluid ``parcels'' that are carried along characteristic curves of the Eulerian velocity is sufficient for deriving the EFE; but it is not necessary.
\medskip

EFE for incompressible inviscid flow were identified as an optimal control system by deriving them using the Hamilton-Pontryagin principle in \cite{BlCrHoMa2000}. This derivation of EFE from an optimal control problem recovered Arnold's interpretation of EFE as geodesic flow on the volume-preserving diffeomorphisms \cite{Ar1966}. It also recovered a standard representation of the Euler equations called the \emph{impulse equations} \cite{Ol1982,Ku1983,Os1989}. Applications of the impulse representation in numerical simulations of EFE were discussed in \cite{RuSm1999}. One disadvantage of the optimal-control formulation of EFE in \cite{BlCrHoMa2000} was that it enforced the pointwise physical constraint that Lagrangian fluid parcels are \emph{frozen} into the flow. That is, the Lagrangian particles were required to follow exactly along the flow lines of the Eulerian velocity vector field. This strict pointwise constraint led to a variational principle whose extremals were not necessarily true minima. Instead, they could have been saddles; so the controllability of Euler fluid flows was left as an open question. Controllability of Euler fluid flows will also be a moot point in the present work. Geometric control \cite{AgSa2004} 
of Euler fluids using Lie methods is reviewed expertly in \cite{AgSa2007}. \smallskip

In the present paper, we first derive EFE from an optimal control problem using the Clebsch variational approach that enforces the frozen-in particle constraint via the back-to-labels, or inverse map, as done previously in \cite{BlCrHoMa2000} for the forward map. This approach produces a symmetric form of EFE analogous to the symmetric, double-bracket form of the $N$-dimensional rigid body dynamics found in \cite{BlCrMaRa2002}. As a result, the Clebsch  representation of the EFE may be written as a coupled system of double-bracket equations in analogy to the corresponding representation of the $SO(N)$ rigid body.   We then re-derive EFE using the method of metamorphosis. From the viewpoint of fluid dynamics, it is interesting that the \emph{same} fundamental EFE appear, even though the metamorphosis approach does not impose the strict requirement that Lagrangian labels follow along characteristic curves of the Eulerian velocity vector field. 

\medskip 

{\bf Plan} The paper has four parts. The first part reviews the properties of the optimal control problem for the $SO(N)$ rigid body \cite{BlCrMaRa2002,BlCr1996,BlBrCr1997}. These properties include a $Q\leftrightarrow P$ exchange-symmetric canonical Hamiltonian formulation that may be rewritten as a coupled system of double-bracket equations on $SO(N)\times SO(N)$. The remarkable features of the $SO(N)$ rigid body provide a model for the paper's subsequent development. The second part shows that an optimization problem for the $SO(N)$ rigid body leads to a derivation of the same dynamical equations as found from its optimal control problem. The third part derives the EFE from an optimal control problem that constrains Lagrangian trajectories to follow Eulerian flow characteristics locally at every point. This approach results in the well-known Clebsch formulation of the EFE \cite{Lamb1932}. In the Clebsch formulation, the same properties of canonical exchange symmetry and double-bracket dynamics emerge as for the $SO(N)$ rigid body. The fourth part derives the EFE from an optimal control problem based on metamorphosis that constrains the Lagrangian trajectories only in an $L^2$ sense. The EFE still reappear from this optimal control problem, albeit with a modified pressure and broken exchange symmetry in the canonical equations. The two differences are immaterial, though, because the term that breaks exchange symmetry in the metamorphosis approach appears in its resulting EFE as merely a redefinition of pressure, i.e., a {\it gauge transformation}, which has no effect at all on the solutions of the Euler fluid equations. 
\smallskip 

{\bf Optimal control of the {\normalsize$N$}-dimensional rigid body} 
Let us recall the standard and symmetric forms of the equations for a rigid body in $N$ dimensions.
We first recall that the left-invariant generalized rigid body equations on
$\operatorname{SO}(N)$ may be written as \cite{Man1976,Ra1980}
\begin{equation}
\dot Q= Q\Omega\,,  \qquad 
\dot M= [M,\Omega]\,, 
\label{rbl}
\end{equation}
where $Q\in \operatorname{SO}(N)$ denotes the configuration space variable (the orientation of the body), $\Omega=Q^{-1}\dot{Q} \in\mathfrak{so}(N)$ is the body angular velocity and
\begin{equation}
M:=J(\Omega)=\Lambda\Omega +\Omega\Lambda \in
\mathfrak{so}(N)^*
\end{equation}
 is the body angular momentum. Here the positive definite operator 
$J: \mathfrak{so}(N) \rightarrow  \mathfrak{so}(N)^* $ is symmetric
with respect to the matrix trace inner product 
\begin{equation}
\langle A,B\rangle = \frac{1}{2}\mbox{tr}(A^TB)
\,.
\label{traceprod}
\end{equation}
The diagonal matrix $\Lambda$ satisfies  $\Lambda_i + \Lambda_j >0$ for
all $i \neq j$. For $n=3$ the elements of $\Lambda_i$
are related to the standard diagonal moment of inertia tensor $I$ by
$I_1 = \Lambda_2 + \Lambda_3$,  $I_2 = \Lambda_3 + \Lambda_1$,
          $I_3 = \Lambda_1 + \Lambda_2$.
          \smallskip

The equation $ \dot{ M } =  [ M, \Omega
] $ is readily checked to be the Euler-Poincar\'e equation on
$\mathfrak{so}(N)^*$ for the Lagrangian
\[
l ( \Omega ) = \frac{1}{2}  \left\langle  \Omega , J
( \Omega )
\right\rangle .
\]

\begin{thm}
The left-invariant rigid body dynamics is given by
the symmetric system of first-order equations
\begin{equation}
\dot Q = Q\Omega
\,, \qquad
\dot P= P\Omega
\,,
\label{rbnl}
\end{equation}
where $\Omega$ is regarded as a function of $Q$ and $P$ via the
equations
\begin{equation}
\Omega :=J^{-1}(M)
\in \mathfrak{so}(N),
\qquad  M := Q^TP-P^TQ.
\label{Om-em}
\end{equation}
\end{thm}
{\bf Sketch of proof:} The result follows by direct substitution of equations (\ref{rbnl}) into (\ref{rbl}), as shown in \cite{BlCrMaRa2002,BlCr1996,BlBrCr1997}. \hfill $\qed$

\begin{remark}
The exchange-symmetric system \emph{(\ref{rbnl})} for $SO(N)$ rigid body motion is canonically Hamiltonian for $H(Q,P)=\frac{1}{4}\langle J^{-1}M(Q,P),M(Q,P)\rangle$ with Poisson bracket $\{Q,P\}=Id$. 

The map $M: T^*SO(N)\to so(N)^*$ in \emph{(\ref{Om-em})} is the cotangent-lift \emph{momentum map} for the (left) action of the Lie group $SO(N)$ on itself. Substituting equations \emph{(\ref{Om-em})} into \emph{(\ref{rbnl})} reformulates them on $SO(N)\times SO(N)$ as a coupled double-bracket system \cite{BlCrMaRa2002,BlCr1996,BlBrCr1997}. 
\end{remark}

The symmetric form of the $n$-dimensional rigid body equations on $SO(N)$  in (\ref{rbnl}) is interesting for our purposes here, because these rigid-body equations can also be derived from the following optimal control problem.

\begin{dfn}[Optimal control problem for the $\operatorname{SO}(N)$ rigid body]
\label{rboptcontprob}$\quad$\\
          Let $T >0 $, and let $Q _0, Q _T \in \operatorname{SO}(N)$
be fixed. The rigid-body optimal control problem is given by
\begin{equation}
\mathop{\rm min}_{U\in
\mathfrak{so}(N)}\frac{1}{4}\int\limits _0^T
\langle U,J(U)\rangle \mbox{d} t
\,,
\label{optr}
\end{equation}
subject to the constraint on $U$ that there be a curve
$Q (t) \in \operatorname{SO}(N)$ such that
\begin{equation}
\dot Q=QU\qquad Q(0)=Q_0,\qquad Q(T)=Q_T
\,.
\label{eqnr}
\end{equation}

\end{dfn}

In the framework of the symmetric representation of the rigid body equations (\ref{rbnl}) the following theorem may be proved.

\begin{thm} [Bloch et al. \cite{BlCrMaRa2002}] 
The rigid body optimal control problem given in Definition \emph{\ref{rboptcontprob}}
has extremal evolution equations \emph{(\ref{rbnl})} where $P$ is the costate vector given by the Pontryagin maximum principle. The optimal control relation in this case is given by
\begin{equation}
U=J^{-1}(Q^TP-P^TQ)
\,.
\label{cntrloptimiz}
\end{equation}
\end{thm}

\begin{dfn}[Optimization problem for the $\operatorname{SO}(N)$ rigid body]\label{rboptimizprob}$\quad$\\
          Let $T >0 $, and let $Q _0, Q _T \in \operatorname{SO}(N)$
be fixed. The rigid-body optimization problem is given by
\begin{equation}
\mathop{\rm min}_{U\in
\mathfrak{so}(N)}\frac{1}{2}\int\limits _0^T
\Big[\,
\langle U,J(U)\rangle 
+\frac{1}{\sigma^2}\|\dot Q-QU\|^2
\, \Big]
\mbox{d} t
\,,
\label{optimcost}
\end{equation}
for a positive real constant $\sigma^2 $, a given metric $\|\,\cdot\,\|: T\operatorname{SO}(N)\to\mathbb{R}$ and subject to the endpoint conditions on $Q$ that the curve $Q (t) \in \operatorname{SO}(N)$ satisfy
\begin{equation}
Q(0)=Q_0,\qquad Q(T)=Q_T
\,.
\label{endpt}
\end{equation}

\end{dfn}

\begin{remark}
The problem statement (\ref{optimcost}) optimizes the cost  (\ref{optr}) of using the controls, for a given tolerance $(\sigma)$ in satisfying the state equations.  
\end{remark}

Suppose the metric $\|\,\cdot\,\|:T\operatorname{SO}(N)\to\mathbb{R}$ in (\ref{optimcost}) is given by
\begin{equation}
\|\,\dot Q-QU\,\|^2={\rm tr}\Big((\dot Q-QU)^T \,K(\dot Q-QU) \Big)
\,,
\quad\hbox{with} \quad K^T=K
\,,
\label{metrix}
\end{equation}
as obtained from the trace inner product for matrices in (\ref{traceprod}), where, as expected for a metric, $K$ is symmetric.
In this framework, the following theorem may be proved.

\begin{thm} [Optimization equations for the $\operatorname{SO}(N)$ rigid body] 
The rigid body optimization problem given in Definition \emph{\ref{rboptimizprob}}
yields extremal evolution equations,
\begin{equation}
K(\dot Q - QU )= \sigma^2P
\,, \qquad
\dot P= -PU^T = PU
\,,
\label{rboptimiz}
\end{equation}
obtained from variations of the cost function {\rm(\ref{optimcost})} in the variables $\dot{Q}$ and $Q$, respectively.
The optimal controls in this case are again given by equation (\ref{cntrloptimiz}).
\end{thm}

{\bf Sketch of proof:} The extremal evolution equations in (\ref{rboptimiz}) follow from stationarity of the cost function {\rm(\ref{optimcost})} under variations in $\dot{Q}$ and $Q$, respectively. Independently requiring stationarity of the cost {\rm(\ref{optimcost})} under variations of the control variable $U$  yields
\begin{equation}
J(U)=\frac12(Q^TP-P^TQ)
\,.
\label{rbloptimz-angmom}
\end{equation}
This reproduces the momentum map in (\ref{Om-em}) and thereby recovers the optimal control relation in (\ref{cntrloptimiz}). \hfill $\qed$

\begin{thm} [$\operatorname{SO}(N)$ rigid body motion optimizes (\ref{optimcost}) ] 
For any value of the tolerance $\sigma^2$, the optimization problem for the $\operatorname{SO}(N)$ rigid body in Definition {\rm\ref{rboptimizprob}} yields an evolution equation in the same form as the angular momentum equations for the rigid body, when written as 
\begin{equation}
\dot M= [M,U]
\,, \quad\hbox{with}\quad
M:=J(U)
\,,
\label{rbloptimz}
\end{equation}
and $J(U)$ is given by the control relation (\ref{rbloptimz-angmom}).
\end{thm}
{\bf Sketch of proof:} The extremal evolution equations (\ref{rboptimiz}) combine with the optimal control relation (\ref{cntrloptimiz}) to produce 
\begin{equation}
\frac{dJ(U)}{dt} = [J(U),U] + \frac{\sigma^2}{2}P^T(K^{-1}-K^{-T})P
\,,
\label{rboptimizJ}
\end{equation}
The last term vanishes, by symmetry of $K$, thereby recovering the motion equation (\ref{rbloptimz}) for any value of $\sigma^2$.  \hfill $\qed$\\

Next, we will derive the symmetric form of the Euler fluid equations on the volume-preserving diffeomorphism group SDiff. These equations correspond to the symmetric  equations (\ref{rbnl}) on $SO(N)$.  We will also compare the equations that result for the control dynamics as they are obtained from the optimal control, and optimization approaches. \medskip
\rem{
Perhaps not unexpectedly, the symmetric form of EFE on the volume preserving diffeomorphisms will be derived from the optimal control problem associated with the exact-matching approach for image registration.
}

{\bf Optimal control for fluids: Clebsch approach}\\
Let us begin by explaining the statement made in the Introduction that metamorphosis penalizes fluid parcels that deviate from their flow lines, but does not force them 
to follow exactly along the flow lines of the fluid velocity. 
We recall that Lagrangian fluid dynamics provides evolution equations for particles moving with a fluid flow. This is typically expressed via the forward flow map $g_t:=g(t)$ from the fluid reference configuration at time $t=0$ (for which $g_0=Id$) to the fluid flow domain $\mathcal{D}$ at a later time $t$. The flow lines are given by $x_t=g_tl$ with $x_0=l$, the vector label. (Here vector and covector indices are understood, but not written explicitly.) The forward map is taken to be a diffeomorphism, so the flow preserves topology. That is, the fluid particles are imagined to be unable to cavitate, superimpose or jump.

For an $n$-dimensional fluid flow, the flow line relation $x_t=g_tl$ of the flow map
$g_t:\,\mathbb{R}^n\times\mathbb{R}\mapsto\mathbb{R}^n$ specifies the spatial position at time $t$ of
the fluid particle that has \emph{label} $l=x_0$. The inverse
map $g_t^{-1}$ is the \emph{back-to-labels map},
\footnote{The back-to-labels map was first named so, and used as a sufficient
variable to describe and analyze the incompressible Euler equations, in \cite{Co2001}},
which gives the label of the particle that occupies
position ${x}$ at time $t$ as the vector function $l_t=g_t^{-1}{x}$.  The Eulerian velocity field
$u_t=u({x},t)$ gives the velocity of the fluid particle that occupies position ${x}$ at time $t$ as
\[
\dot{g}_t = u_t \circ g_t
\quad\hbox{or}\quad 
{\dot{x}}({l},t)={u}({x}({l},t),t).
\]
The vector components of particles whose labels $l_t$ are \emph{frozen} into an ideal fluid flow each satisfy the \emph{advection law} obtained from the time derivative (tangent) of the back-to-labels map 
$l_t=g_t^{-1}{x}=l(t,x)$,
\begin{equation}
\label{label eqn}
\frac{D l}{Dt}: = \pa_tl + u\cdot\nabla{l} = 0.
\end{equation}
The quantity $v=-\pa_tl(t,x) =u\cdot\nabla{l} $ is called the \emph{convective velocity} \cite{HoMaRa1986}. Satisfying $v={\rm Ad}_{g^{-1}}u$, the convective fluid velocity $v$ is to the Eulerian spatial fluid velocity $u$ as body angular velocity is to spatial angular velocity for a rigid body
\footnote{The relation $v={\rm Ad}_{g^{-1}}u$ between the spatial fluid velocity $u$ and the convective fluid velocity $v$ could also be used as either a constraint, or a penalty. This would be an alternative state equation and, thus, an alternative approach to the present considerations.}.
\smallskip

The optimal control problem for EFE may be posed using the vector $L^2$ pairing $\langle\,a\,,\, b\,\rangle=\int_{\mathcal{D}} a^T\,b\,{\rm d}^nx$ as
\begin{equation}
\label{OC prob}
\min_{u_t(\cdot)}\int\limits ^T_0 \ell(u)\, \mbox{d} t
\quad\hbox{with}\quad
\ell(u) = \frac12\langle u,u\rangle
,
\end{equation}
subject to
$
\operatorname{div}u=0
$,
$
\partial_t{g}=u\circ g
$, 
with 
$
g(l,0)=g_0(l)
,\
g(l,T)=g_T(l)
$
fixed and, for flow in all of space, suitable conditions at infinity. 
\smallskip

The optimal control problem (\ref{OC prob}) is of course identical to the standard Hamilton principle
for ideal fluid mechanics and it has been solved previously in terms of the forward map \cite{BlCrHoMa2000}. However, that approach using the forward map did \emph{not} produce the symmetric form of the EFE. Here, we solve problem (\ref{OC prob}) and find the symmetric form of the EFE by introducing Lagrange multipliers $({\pi},\,k)$ into the kinetic-energy cost that impose incompressibility and the pointwise tangent relation (\ref{label eqn}) along the \emph{inverse flow}, i.e.,  the back-to-labels map $l(t,x)$,
\begin{equation}
S(u,l,{\pi},k)=\int\limits _{0}^{T}\!\!
\Big( \ell(u) 
+
\underbrace{\,
\langle {\pi},\pa_tl + u\cdot\nabla{l}\rangle\,
}_{\hbox{Constraint}}
-
\underbrace{\,
\langle k,\operatorname{div}u\rangle \,
}_{\hbox{Constraint}}
\!\Big)\mbox{d} t
.
\label{Lag1efe}
\end{equation}
The problem may be recast as:
$\min S$, subject to the two constraints $\operatorname{div}u=0$ and $\pa_tl + u\cdot\nabla{l}=0$, as well as spatial boundary conditions and endpoint conditions that $l(0,x)$ and $l(T,x)$ are fixed. 

\begin{remark}
The absence of weak solutions in $L^2$ for the Euler equations is a well-known problem, both for the initial value problem and for the geodesic minimization problem, see, e.g.,  \cite{MjBe2002}. 
However, for the stronger norms used in typical applications of metamorphosis in image analysis \cite{HoTrYo2009} and with the $H^1$ norm for the averaged Euler equations \cite{MaRaSh2000}, one may assume that all entities exist, functions are smooth and integrals are finite. Consequently, we will perform the calculations below with a sufficiently smooth Lagrangian $\ell(u)$ that these assumptions hold in either the optimal control problem in equation (\ref{Lag1efe}) or the optimization problem in equation (\ref{Lag2efe}). We will then specialize the results to the case (\ref{OC prob}) of $u\in L^2$ when referring to the Euler equations, knowing that the result is only formal in this case. 
\end{remark}

For the optimal control problem in equation (\ref{Lag1efe}), the following results may be proved for a sufficiently smooth Lagrangian $\ell(u)$.


\begin{thm}\label{T:2}
The extremals of $S(u,l,{\pi},k)$ are given by
\begin{align}\label{OCextremals}
&
 \frac{\delta \ell }{\delta u}
 + 
 {\pi}\nabla{l}+\mbox{\rm grad}\,k=0
 \,,\quad \frac{D {\pi}}{Dt}=0=\frac{D l}{Dt} 
 \,,\\
&
\hbox{and}\quad 
\mbox{\rm div}\,u=0
\quad\hbox{with}\quad 
\frac{D }{Dt} := \pa_t + u\cdot\nabla
.
\nonumber
\end{align}
\end{thm}
\textbf{Sketch of proof:}  The variations of $S(u,g,{\pi},k)$ yield
\begin{eqnarray*}
&&\delta S
=
\int\limits ^T_0
\bigg(
\bigg\langle  \frac{\delta \ell }{\delta u}
 + 
 {\pi}\nabla{l}+\mbox{\rm grad}\,k ,\delta u \bigg\rangle
 - \langle \delta  k,\mbox{ div}\, u\rangle
 \\
&& 
+\
 \langle \delta{\pi},\pa_tl + u\cdot\nabla{l}\rangle
- \langle \pa_t{\pi} + \mbox{div}\, {\pi} u, \delta l \rangle
\bigg)\mbox{d} t
+\
\Big[{\pi} \,\delta l\Big]_0^T
\!.
\end{eqnarray*}
System (\ref{OCextremals}) follows immediately, upon using div$\,u=0$ 
and noting that $\delta u(\infty,t)=\delta l (0,x) = \delta l (T,x) =0$. \hfill $\qed$

\begin{remark}
The $({\pi},l)$ equations \emph{(\ref{OCextremals})} are symmetric under exchanging ${\pi}\leftrightarrow l$. These are the analogs for  {\rm SDiff} of the $P \leftrightarrow  Q$ symmetric equations \emph{(\ref{rbnl})} for $SO(N)$. The representation of the control variable $u\in \mathfrak{X}$ (a divergenceless vector field) in terms of the state variables $\pi,l,k$ is given by inverting the map $m(u)\in \mathfrak{X}^*$ in
\begin{equation}
m(u) := \frac{\delta \ell }{\delta u} = -\,  {\pi}\nabla{l}-\nabla{k} 
\,,
\label{momapR}
\end{equation}
obtained from the variations in $u$ of the cost function $S$. 
The Clebsch map $m: T^*{\rm SDiff}\to \mathfrak{X}^*$ in \emph{(\ref{momapR})} is the cotangent-lift \emph{momentum map} for the (right) action of ${\rm SDiff}$ on itself.
\end{remark}

\begin{thm}
Substitution of the Lagrangian $\ell(u)$ in \emph{(\ref{OC prob})} into system \emph{(\ref{OCextremals})} for the extremals of $S$ yields EFE. 
\end{thm}

\textbf{Sketch of proof:}  System (\ref{OCextremals}) implies the following version of Kelvin's circulation theorem for ideal fluids
\begin{eqnarray}
\frac{d }{dt} \oint_{c(u)} \frac{\delta \ell }{\delta u}\cdot \mbox{d} x
&=&
 \oint_{c(u)} \bigg[ \frac{D }{D t} \frac{\delta \ell }{\delta u }
 + (\nabla u)^T\cdot \frac{\delta \ell }{\delta u} \bigg] \cdot \mbox{d} x
\nonumber\\
&&\hspace{-8mm}
=
- \oint_{c(u)}
\bigg(
 \frac{D{\pi} }{Dt} \mbox{d} l + {\pi} \mbox{d}  \frac{Dl }{Dt} + \mbox{d}\frac{Dk}{Dt}
 \bigg)
 =
 0
 \,,
 \label{Kel-circ-thm}
\end{eqnarray}
whose RHS vanishes upon using (\ref{OCextremals}) and noting that the integral of an exact form vanishes when taken around the closed loop $c(u)$ moving with the Eulerian velocity $u$. For the Euler case (\ref{OC prob}), in which 
$\delta \ell/\delta u=u$, this calculation recovers EFE as
\begin{equation}\label{EFE-velocity}
\frac{D u}{D t} = -\, \nabla \bigg(\frac{Dk}{Dt} + \frac{u^2}{2} \bigg) =: -\, \nabla p
\quad\hbox{and} \quad
\mbox{div}\,u=0
\,.
\qquad \qed
\nonumber
\end{equation}
{\bf Vorticity dynamics}  By taking the curl, the Euler equations for the vorticity 
$w=\mbox{curl}\,u$ are found to be
\begin{equation}\label{EFE-vorticity}
\frac{\partial w}{\partial t}=[w,u]
, \quad
w=\nabla{\pi}\times\nabla{l}
,\quad
\operatorname{div} u=0,
\end{equation}
where $[w,u]= w\cdot\nabla u - u\cdot\nabla w$ is the Lie bracket of the divergenceless vector fields $\mathfrak{X}$ on the flow domain $\mathcal{D}$. The interpretation of (\ref{EFE-vorticity}) is that the symplectic 2-form $d\pi\wedge dl$ is frozen into the forward map.

\begin{remark}
When $\delta \ell/\delta u=u$ the system  \emph{(\ref{OCextremals})} for the extremals of $S$ recovers the classical Clebsch representation of ideal incompressible fluid flow \cite{Lamb1932}. 
The vorticity equation \emph{(\ref{EFE-vorticity})} is canonically Hamiltonian for 
\[
H({\pi},l)=\frac{1}{2}\langle (\Delta)^{-1}w({\pi},l),w({\pi},l)\rangle
\,,
\]
with Poisson bracket $\{{\pi}(x),l(x')\}=-\delta(x-x')$. The map \emph{$w: T^*{\rm SDiff}\to \mathfrak{X}^*$} on $\mathcal{D}$ is the cotangent-lift momentum map for the action of \emph{SDiff} on itself. Substituting equations \emph{(\ref{Om-em})} into \emph{(\ref{rbnl})} reformulates them on \emph{SDiff $\times$ SDiff} as a coupled double-bracket system \cite{GBRa2009}.  
\end{remark}

{\bf Metamorphosis  and the optimization problem for fluids}\\
Now we come to the point of formulating the optimization problem (\ref{OC prob}) for EFE whose extremals are sure to be minima. This is the metamorphosis formulation, which solves  
problem (\ref{OC prob}) by introducing an additional norm into the kinetic-energy cost that imposes incompressibility as a constraint and treats the tangent relation (\ref{label eqn}) for the back-to-labels map as merely an optimization penalty. This is 
\begin{equation}
S(u,l,k)=
\!\!
\int\limits ^T_0\!\!
\Big( \ell(u) 
+ \frac{1}{2\sigma^2} \underbrace{\
\| \pa_tl + u\cdot\nabla{l}\|_{L^2}^2\
}_{\hbox{Penalty}}
-
\underbrace{\,
\langle k,\operatorname{div}u\rangle\, 
}_{\hbox{Constraint}}
\Big)\mbox{d} t
\,.
\label{Lag2efe}
\end{equation}
The problem may be recast as:
$\min S$, subject to $\operatorname{div}u=0$, spatial boundary conditions, endpoint conditions that $l(0,x)$ and $l(T,x)$ are fixed and penalize for the error in the $L^2$ norm $\|\pa_tl + u\cdot\nabla{l}\|_{L^2}^2$. For $\sigma^2>0$, when extremals exist they will be minima. 

For this problem, the following results may be proved.

\begin{thm}\label{T:2}
The extremals of $S(u,l,k)$ are given by
\begin{align}\label{OCminima}
&
 \frac{\delta \ell }{\delta u}
 + 
 {\pi}\nabla{l}+\mbox{\rm grad}\,k=0
 \,,\quad
 \frac{D {\pi}}{Dt}=0
 \,, \quad
  \frac{D l}{Dt} = \sigma^2{\pi}
 \,,\\
&
\hbox{and}\quad 
\mbox{\rm div}\,u=0
\quad\hbox{with}\quad 
\frac{D }{Dt} := \pa_t + u\cdot\nabla
.
\nonumber
\end{align}
\end{thm}
\textbf{Sketch of proof:}  After defining the canonical momentum by $\sigma^2{\pi}=\pa_t l + u\cdot\nabla l$ obtained from varying $ \pa_t l $, the other variations of $S(u,l,k)$ yield
\begin{eqnarray*}
\delta S
&=&
\int\limits ^T_0
\bigg(
\bigg\langle  \frac{\delta \ell }{\delta u}
 + 
 {\pi}\nabla{l}+\mbox{\rm grad}\,k ,\delta u \bigg\rangle
 - \langle \delta  k,\mbox{ div}\, u\rangle
 \\
&& \hspace{2cm}
-\
 \langle \pa_t{\pi} + \mbox{div}\, {\pi} u, \delta l \rangle
\bigg)\mbox{d} t
+\
\Big[{\pi} \, \delta l\Big]_0^T.
\end{eqnarray*}
System (\ref{OCminima}) follows immediately, upon using div$\,u=0$ 
and noting that $\delta u(\infty,t)=\delta l (0,x) = \delta l (T,x) =0$. \hfill $\qed$

\begin{remark}
Equations \emph{(\ref{OCminima})} break the exchange symmetry ${\pi}\leftrightarrow l$, which is only restored in the limit $\sigma^2\to0$.
\end{remark}

\begin{thm}
Substitution of the Lagrangian $\ell(u)$ in \emph{(\ref{OC prob})} into system \emph{(\ref{OCminima})} for the minima of $S$ again yields EFE,  for \emph{any value} of $\sigma^2>0$. 
\end{thm}

\textbf{Sketch of proof:}  System (\ref{OCminima}) implies the following version of Kelvin's circulation theorem
\begin{eqnarray*}
\frac{d }{dt} \oint_{c(u)} \frac{\delta \ell }{\delta u}\cdot \mbox{d} x
&=&
 \oint_{c(u)} \bigg[ \frac{D }{D t} \frac{\delta \ell }{\delta u }
 + (\nabla u)^T\cdot \frac{\delta \ell }{\delta u} \bigg] \cdot \mbox{d} x
\\
&&\hspace{-4mm}
=
- \oint_{c(u)}
\bigg(
 \frac{D{\pi} }{Dt} \mbox{d} l + {\pi}\, \mbox{d}  \frac{Dl }{Dt} + \mbox{d} \frac{Dk}{Dt}
 \bigg)
 =
 0
 \,,
\end{eqnarray*}
whose RHS vanishes upon using (\ref{OCminima}) and noting that ${\pi} \mbox{d}(Dl/Dt) = \mbox{d}(\sigma^2{\pi}^2/2)$ is exact, so it does not contribute to the integral taken around the closed loop $c(u)$ moving with the Eulerian velocity $u$. For the case $\delta \ell/\delta u=u$, this calculation recovers EFE as
\[
\frac{D u}{D t} = -\, \nabla \bigg(\frac{Dk}{Dt} + \frac{u^2}{2} + \sigma^2\frac{{\pi}^2}{2} \bigg) =: -\, \nabla p
\,,
\]
and div$\,u=0$ yields pressure $p$ \emph{independently} of $\sigma^2$.  \hfill $\qed$

\begin{remark}
One concludes that an optimal control problem and an optimization problem for incompressible ideal fluid flow both yield the \emph {same} geodesic Euler fluid equations for their control relations, although they represent \emph{different} Lagrangian dynamics. 
The EFE equations derived from either the optimal control problem based on the Clebsch approach, or the metamorphosis optimization approach are equivalent up to a \emph{gauge tranformation} of the pressure, which is immaterial for incompressible flow, since the pressure in this case is determined independently by preservation of the constraint \emph{div}$\,u=0$. The emergence of equivalent EFE from either optimal control or  optimization means that the forward map in the Lagrangian picture of ideal fluid dynamics implies the Eulerian picture, but not vice versa. 

The label paths defined by the inverse map $l_t=g_t^{-1}{x}$ may ``forget their way'' and their dynamics may diverge from the requirements of the Eulerian velocity characteristics as in (\ref{OCminima}) for \emph{any finite value} of $\sigma^2\ge0$ without changing the optimal outcome. Taking the limit of sigma to zero yields exactly the same relations for the label dynamics of both the optimization problem and the optimal control problem. However, this result is immaterial to the emergence of the EFE, because both problems yield the \emph {same} Euler fluid equations, even though their Lagrangian parcel dynamics are \emph{different}. The result arises from the \emph{gauge freedom} in the definition of the fluid pressure for an incompressible flow. One notes that the gauge freedom in defining the pressure for incompressible fluid flow is independent of the symmetry of fluid dynamics under relabeling of their Lagrangian coordinates. The relabeling symmetry implies the conservation of circulation in (\ref{Kel-circ-thm}) and holds for both Lagrangian cost functions (\ref{Lag1efe}) and (\ref{Lag2efe}), independently of the definition of pressure. If the relabeling symmetry were broken to a subgroup corresponding to invariance of fluid properties appearing in the thermodynamic definition of pressure, the two formulations may produce a nontrivial difference. This feature will be investigated elsewhere.

\end{remark}

{\bf Acknowledgements} We are grateful to A. M. Bloch, P. Constantin, 
C. J. Cotter, B. A. Khesin, J. E. Marsden, J. Monaghan and T. S. Ratiu for many useful and pleasant conversations about these and related matters. This work was partially supported by a Wolfson Award from the Royal Society of London.

\bibliography{EulControl-1}
\bibliographystyle{apsper}


\end{document}